\documentclass[%
,twocolumn
 ,secnumarabic%
,amssymb, amsmath,nobibnotes,superscriptaddress,aps,prl]{revtex4}
\usepackage{bm}%
\usepackage{graphicx}
\usepackage{graphics}
\usepackage{xspace}

\newcommand{\CO}{CO$_2$\xspace}

\begin{document}
\title{New insights on the high-pressure phase diagram of molecular \CO}%

\author{Valentina M. Giordano}%
\altaddress{Present address: ESRF, 6 rue Jules Horowitz, BP220, 38043 Grenoble CEDEX, France}
 \affiliation{Physique des Milieux Denses, IMPMC, CNRS UMR 7590, Universit\'e
Pierre et Marie Curie-Paris VI, 140 rue de Lourmel, 75015 Paris, France}
 \affiliation{LENS, via N. Carrara 1, Polo Scientifico, 50019 Sesto Fiorentino (FI), Italy}
\author{Fr\'ed\'eric Datchi}
 \affiliation{Physique des Milieux Denses, IMPMC, CNRS UMR 7590, Universit\'e
Pierre et Marie Curie-Paris VI, 140 rue de Lourmel, 75015 Paris, France}

\date{\today}%
\begin{abstract}
We report the discovery of a new molecular phase of carbon dioxide at high-pressure and high-temperature. Using
x-ray diffraction, we identify this phase as the theoretically predicted high-temperature $Cmca$ phase [Bonev
\emph{et al.}, Phys. Rev. Lett., \textbf{91}, 065501 (2003)]. Its relation with phase III, on one hand, and its
relative stability with respect to phase IV, on the other hand, are discussed based on spectroscopic and melting
data. The existence of this strictly molecular phase challenges the interpretation of phases IV  and II as
intermediate phases between the molecular and covalent-bonded forms of \CO.

\end{abstract}
\pacs{} \keywords{} \maketitle

In the last years, the intensive study of simple molecular systems (H$_2$, N$_2$, CO, CO$_2$ \textit{etc.}) at
very high pressures ($P$) and temperatures ($T$) have led to a wealth of remarkable discoveries, including new
solid structures, insulator-metal transitions, symmetrization of H-bonds, polymerization or the formation of new
covalent bonds. These findings have had important repercussions both on the fundamental level, since their
understanding is usually challenging, and from a technological viewpoint, since they may open, for example, new
routes for energy storage.

\begin{figure}
\includegraphics[width=6.5cm]{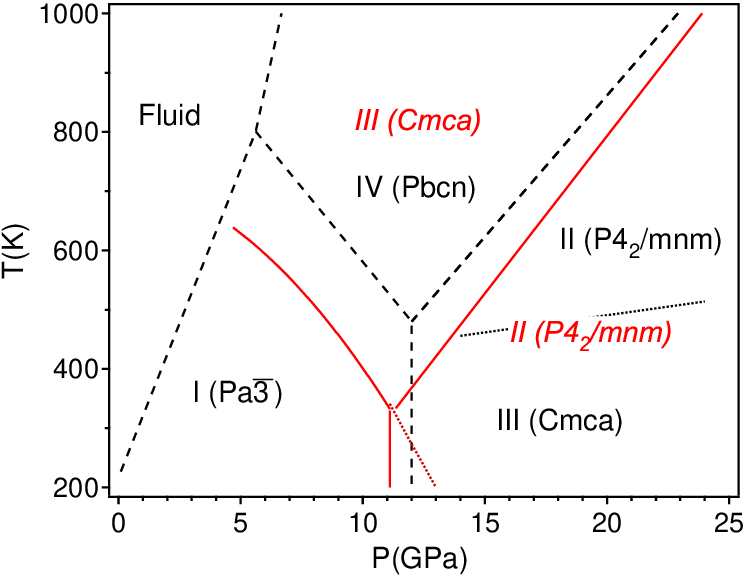}
\caption{(Color online) Experimental~\cite{IotYoo2001} (black dashed lines) and theoretical~\cite{BonGyg2003}
(red solid lines) phase diagram of carbon dioxide. The theoretically predicted stable phases are indicated in
italics, and the proposed experimental ones, in upright characters. The dotted lines are speculated kinetic
lines.} \label{Fig0}
\end{figure}

In this context, the discovery in 1999 by Iota \emph{et al.}~\cite{IotYooCyn1999} of a non-molecular, quartz-like, phase of
carbon dioxide by heating above 50 GPa has driven a large attention onto the behavior of this compound at
elevated pressures. This phase was described as an extended network of four-fold coordinated carbon atoms, with a
unusually large bulk modulus. Subsequently, new molecular phases were
uncovered, named II and IV (Fig.~\ref{Fig0})~\cite{IotYoo2001,Yoo2001}. From the large values of the C$=$O bond lengths found for these two phases, and the bent geometry of the molecule in phase IV, they were presented as precursors to the non-molecular phase. This interpretation, however, is not supported by a theoretical study by Bonev and co-authors~\cite{BonGyg2003} who found, first, that the equilibrium C$=$O bond length in any of the proposed structures is comparable to the free molecule value; and second, the stable structure in the domain of phase IV is not $Pbcn$ with bent molecules, but a molecular $Cmca$ structure with linear molecules.

This controversy enlightens a peculiarity of carbon dioxide compared to other simple molecular systems, that is,
the presence of large metastabilities when going from one solid phase to the other. For example, the transition
between phase I and III is a very sluggish one, spreading over 10 GPa at room
temperature~\cite{YooCyn1999,IotYoo2001,GorGio2004}, and seemingly taking place through an intermediate
structure~\cite{OliJep1998}. These metastabilities make the task of determining the phase lines with accuracy a
difficult one and more importantly, raise the question of which structure is actually the stable one at high P-T.

In this Letter, we report a new investigation of the phase diagram in the pressure region below 20 GPa and for
temperatures up to 950 K. It reveals the presence of a previously unobserved phase (noted \CO-VII) above 640 K, intermediate between phases I and IV. Using x-ray diffraction, we determine the structure of this phase and show that it is the theoretically predicted high P-T $Cmca$ phase. The distinction between \CO-VII and \CO-III is revealed by comparing their Raman spectrum. We also report measurements of the melting lines of this new phase and phase IV which allow us to determine their relative stability. Finally we show that the discovery of this phase has important consequences on the nature of phases IV and II.

\begin{figure}
\includegraphics[width=7cm,clip]{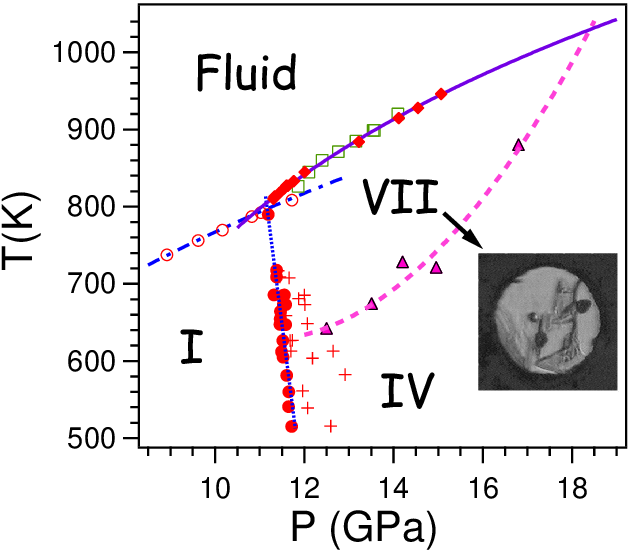}
\caption{(Color online) Presently determined phase transition points. The melting curve of phase I reported in
Ref. \onlinecite{GioDat2006} is shown as the dot-dashed line; $\circ$: melting points of phase I (present work);
$+$ and $\bullet$: I-IV transition measured respectively on loading and unloading; $\blacktriangle$: VII to IV
transition; $\square$: melting points of phase VII; $\blacklozenge$: melting points of phase IV. The experimental
uncertainties on $P$ and $T$ are within the symbol sizes. The curves are fit to the data. The melting points of
phase IV and VII are well fitted by the Simon-Glatzel law~\cite{SG1929} : $T=T_0[1+(P-P_0)/a]^{1/c}$ with
$T_0=805$~K, $P_0=11.15$~GPa, a=46(8)~GPa and c=3.8(6).  The photograph shows the sample of phase VII (with
pressure sensors) viewed through the diamond anvils at 770~K and 11.8 GPa. } \label{Fig1}
\end{figure}

The experiments reported here were done in resistively heated diamond anvil cells. The techniques are identical to
the ones presented in details in Ref. \onlinecite{GioDat2006}.

We first investigated the I-IV transition line between 500 and 790~K by visual observation and Raman
spectroscopy, tuning the pressure around the transition in order to observe it both on loading and unloading. As
Fig.~\ref{Fig1} shows, the unloading transition points lie on a well defined nearly vertical line, whereas the
loading ones are quite scattered. This hysteresis decreases with temperature and nearly disappears above 700~K.
From the unloading line and the previously reported melting curve~\cite{GioDat2006}, we localize the I-IV-Fluid
triple point at 800(5)~K and 11.2(1)~GPa.

\begin{figure}
\includegraphics[width=7cm]{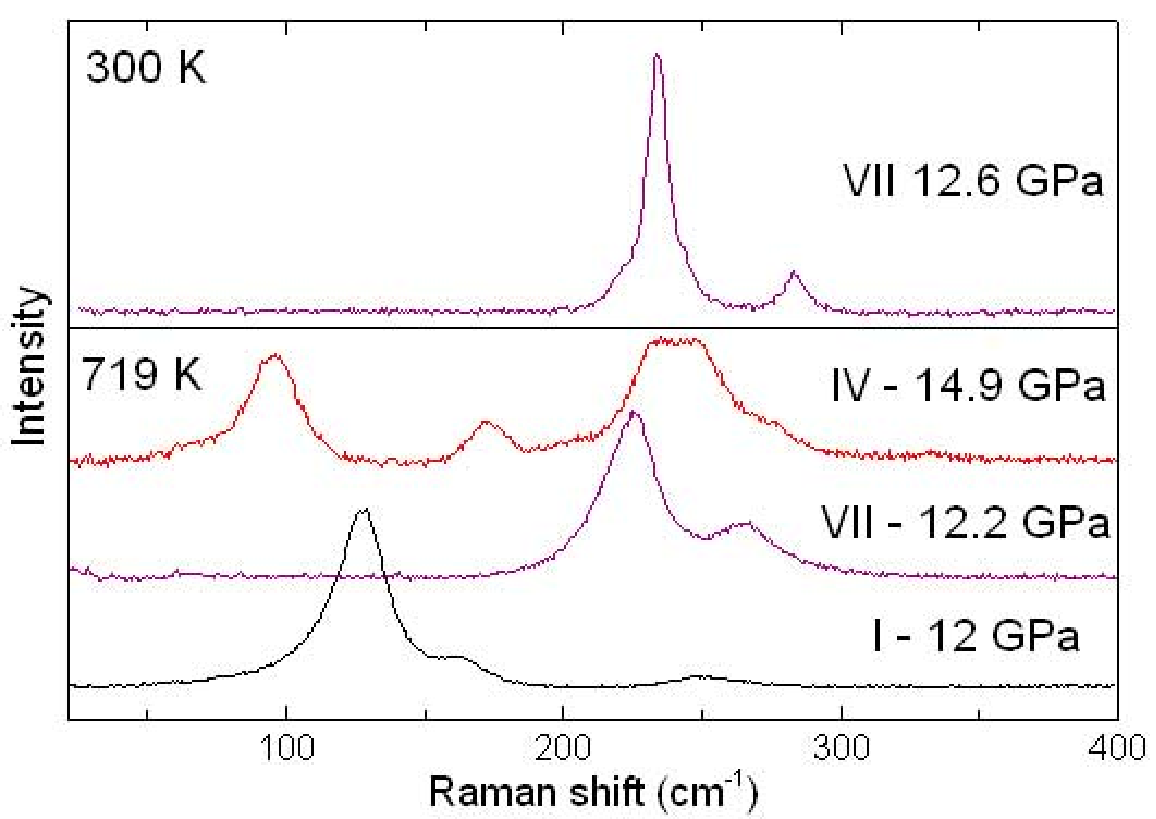}
\caption{(Color online) Raman spectra collected on quenched phase VII at 300~K(top) and along an isotherm at
719~K going from phase I to IV through VII (bottom).} \label{Fig2}
\end{figure}

For $T>640$~K and $P>12$~GPa, compression of phase I systematically produced
a new phase, noted hereafter \CO-VII, which in turn transformed to phase IV by further compression. \CO-VII was
easily identifiable both visually and from its Raman spectrum. The transition was usually sharp and the new phase
visibly birefringent (Fig.~\ref{Fig1}). Fig.~\ref{Fig2} reports the Raman spectra collected along an isotherm at
719~K, following the transition from phase I to IV through phase VII. The characteristics of the Raman
spectrum of \CO-VII will be analyzed below. At 640~K the I-VII and VII-IV transitions took place at 12.3~GPa
and 12.55~GPa respectively. Increasing the temperature, the pressure domain in which phase VII is observed grows
larger: at 726~K the two transitions were localized at 12.1 and 14.2~GPa respectively. Decompression of phase VII
at 705~K led to phase I at 11.8~GPa, i.e. slightly above the IV to I transition pressure (11.4 GPa).

To investigate the structure of phase VII, we collected x-ray diffraction spectra following an isotherm  at 726~K, using the monochromatic ($\lambda=0.3738$~\AA) angular dispersive setup on beamline ID27 of the European Synchrotron Radiation Facility in Grenoble.
The images were collected during rotation of the cell around the $\omega$ axis. The observed $d$-spacings match very well an orthorhombic unit cell with 4 molecules and the following lattice parameters at 12.1~GPa, 726~K: $a=4.313(1)$ \AA, $b=4.746(1)$ \AA, $c=5.948(1)$ \AA.  All the peaks could be indexed by considering three distinct diffracting crystallites rotated around the $c$ axis with respect to each other. Visible peaks obey the following conditions: $hkl\,:\, h+k=2n$; $hk0\,:\,h,k=2n$; $0kl\,:\,k,l=2n$, from which we deduce the space group $Ccmb$. This is the same group as $Cmca$ but with a different choice of axes, i.e. with $a$ and $b$ axes inverted.

\begin{figure}
\includegraphics[width=6.5cm]{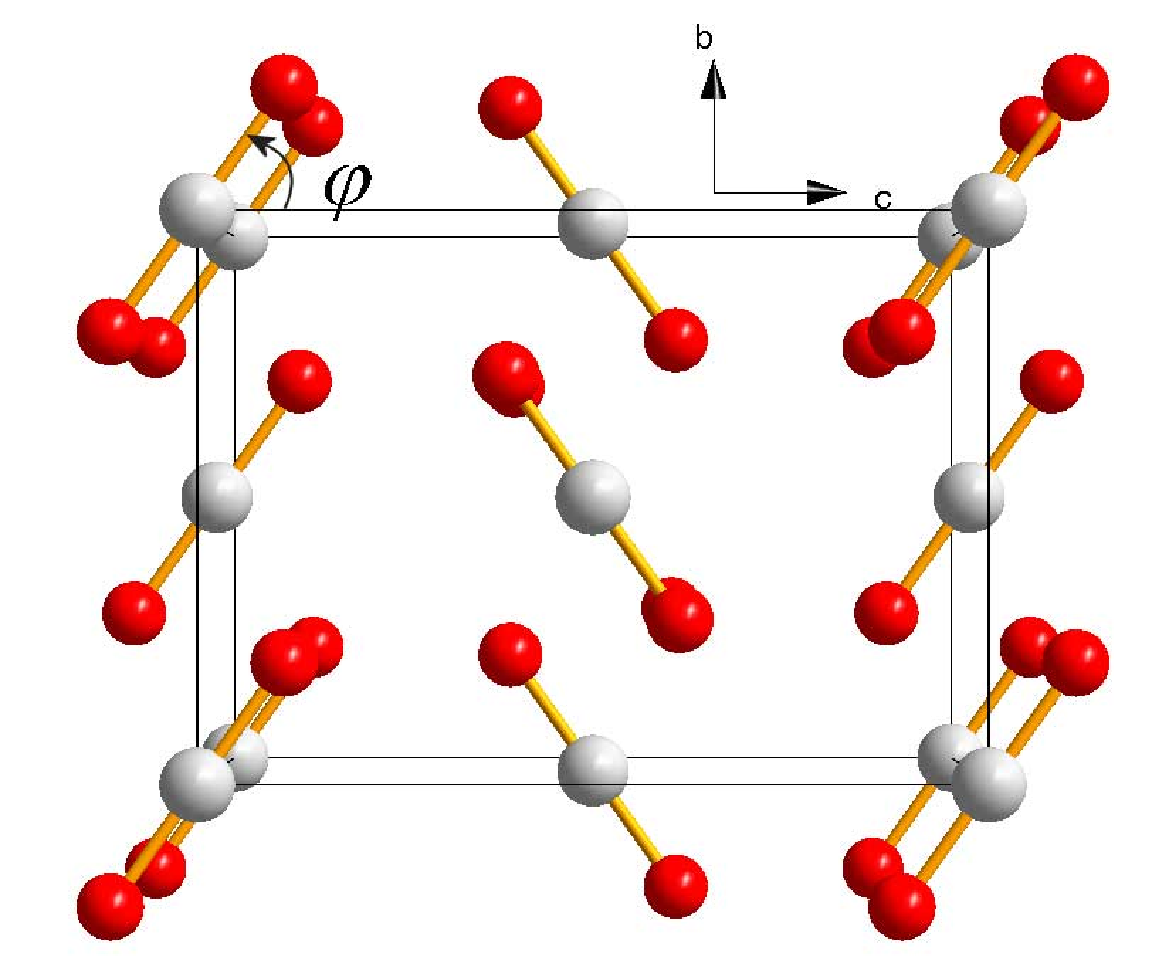}
\includegraphics[width=6cm]{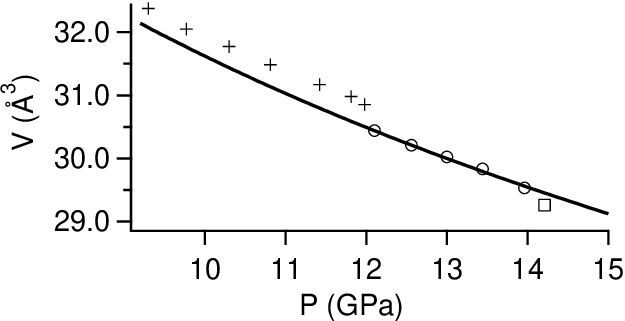}
\caption{(Color online) Structure of the $Cmca$ phase VII at 726~K and 12.1~GPa, projected along the $a$ axis
. The molecules lie on the $(bc)$ plane, tilted by
$\varphi=54.5^\circ$ with respect to the $c$ axis; in the bottom, the volume per molecule vs. pressure is shown
along the isotherm at 726~K for phase I ($+$), VII ($\circ$) and IV ($\square$). The solid line is the
theoretically computed equation of state for the $Cmca$ phase at 0~K~\cite{BonGyg2003}. The volume discontinuity
is $1.2$\% between phase I and VII and $0.66$\% between phase VII and IV.} \label{Fig3}
\end{figure}

The existence of a $Cmca$ polymorph in high-pressure \CO was first predicted by Kuchta and Etters using lattice
dynamics calculations~\cite{KucEtt1988}. In their structure, the axes are in the order $a<b<c$ and the molecules
lie on the $(bc)$ plane, tilted by $\varphi=52^\circ$ with respect to the $c$ axis. More recent calculations
based on density functional theory (DFT)~\cite{Gygi1998,BonGyg2003} found that the optimized $Cmca$ structure,
stable at high $P-T$, was such that $a>b$  and $\varphi=54^\circ$. Since we find the same axis order, we used
this structure as a starting point for refinement. The latter was done against the measured intensities of 16
independent reflections originating from the same crystallite~\footnote{We used the \emph{XDS} [W.J. Kabsch,
Appl. Cryst. 26, 795-800 (1993)] and \emph{Crystals} [P.W Betteridge \emph{et al.} , J. Appl. Cryst. 36, 1487
(2003)] softwares.}. Convergence was obtained down to an $R$-factor of 0.069, yielding for atomic positions, $C$
on (0,0,0) and $O$ on [0,0.212(3),0.109(1)], with isotropic temperature factors $U$ of $0.0637$~\AA$^2$ ($C$) and
$0.0314$~\AA$^2$ ($O$). The refined structure, shown on Fig.~\ref{Fig3}, presents a $\varphi$ angle of
$54.5(1)^\circ$, in excellent agreement with the DFT calculations~\cite{Gygi1998,BonGyg2003}. The C$=$O bond
length is $1.13(1)$~\AA, i.e. slightly shorter than the one found~\cite{DowSom1998} for phase I at 1 GPa
[1.1486(9)~\AA], and again in good agreement with calculations. We can thus conclude that phase VII \emph{is} the
$Cmca$ phase predicted by Bonev \textit{et al.}~\cite{BonGyg2003} to be stable in this $P-T$ range. This is
further corroborated by the excellent agreement observed between the presently determined and calculated equation
of state~\cite{BonGyg2003} (Fig.~\ref{Fig3}, bottom).

\begin{figure}
\includegraphics[width=6.5cm]{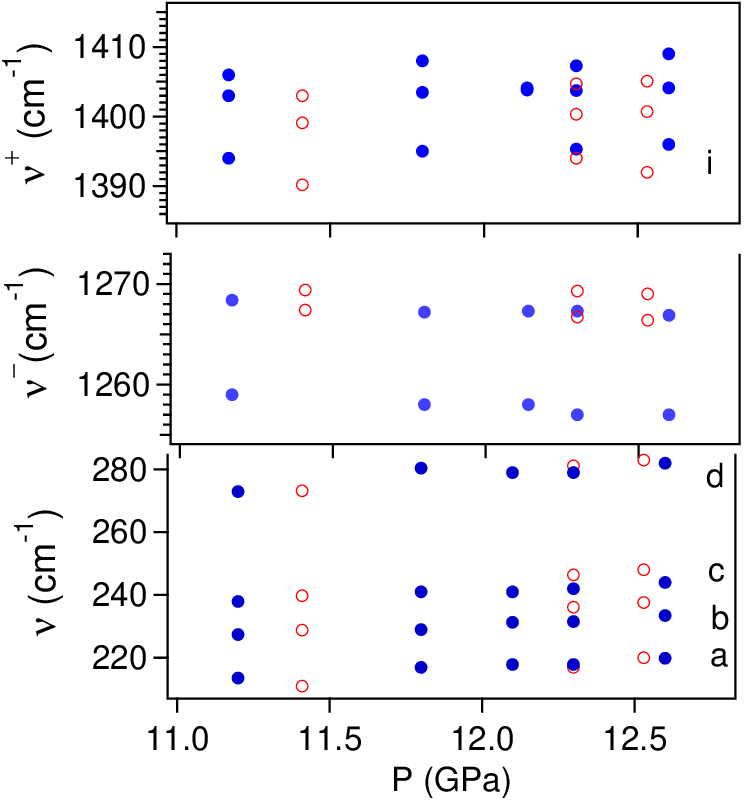}
\caption{(Color online) Room temperature pressure evolution of external and internal modes in phase VII ($\bullet$) and III ($\circ$). The mode i comes from the $\approx1$\% of $^{13}$C$^{16}$O$_2$ molecules.}
\label{Fig4}
\end{figure}

 It is worth pointing out that the $Cmca$ space
group is also the one designated for the room-temperature high-pressure phase III. The structure of this phase
was identified by Aoki \emph{et al.}~\cite{AokYam1994}, and later by Yoo \emph{et al.}~\cite{YooCyn1999}, as the
one predicted by Kuchta and Etters~\cite{KucEtt1988}. However we note that collecting good diffraction data on
phase III has been so far challenging due to the highly strained nature of this phase. As a result, its structure
is still not well constrained. In their work, Bonev \emph{et al.}~\cite{BonGyg2003}, like previous authors,
assimilated the theoretical $Cmca$ phase with phase III, only relocating its stability domain to that of phase IV
(Fig.~\ref{Fig0}) (the $Cmca$ structure was found unstable at 300 K, so phase III was presented as metastable at
this temperature). This raises the question of whether phase VII is identical to phase III. First of all, we note
that the experimental existence domains of the two phases are not connected in the $P-T$ space : phase III
transits to II at $\approx450$~K whereas \CO-VII is only observed above 640~K. To go further, we have compared
their Raman spectra collected during decompression at 300~K. Phase VII could be quenched to room temperature,
which is a common property of all the high $P$-$T$ phases of carbon dioxide. Phase III was produced by
compression of phase I at 300~K and annealed to 440~K to release the stress. The pressure evolution of the Raman
modes is shown in Fig.~\ref{Fig4}.
The four lattice phonon modes and two components of $\nu^+$ and $\nu^-$~\footnote{$\nu^+$ and $\nu^-$ come from the
Fermi resonance between the overtone of the bending 2$\nu_2$ and the symmetric stretching mode $\nu_1$.}, observed
in both cases, match group theory predictions for the $Cmca$ structure.  A
$\approx5$~cm$^{-1}$ frequency difference between the two solids is registered in the central lattice modes (b
and c in the figure) and in the $\nu^+$ region, whereas the difference increases up to 10~cm$^{-1}$  for the
components of $\nu^{-}$. This indicates that the two structures are indeed different. Further work is
however needed to clarify this question.

By contrast with the theoretical phase diagram in Fig.~\ref{Fig0}, the $Cmca$ phase is not the only one present at high temperatures, since it transits to
phase IV at higher pressure. As a matter of fact, we could only obtain phase VII by compression of phase I and
never by heating or decompressing phase IV.   This is reminiscent of the fact that phase III has itself only been
obtained from compression of phase I. Since this was taken as an indication of the metastable character of phase
III, this raises the question of which of phase VII or IV is the stable structure beyond the I-(IV,VII)
transition line.

To answer this,  we have measured the melting curve of \CO for $T > 800$~K, extending the determination presented
in Ref.~\onlinecite{GioDat2006}.  A melting point was defined as the $P-T$ conditions where the solid/fluid
equilibrium was visually observed and the solid phase was identified by its Raman spectrum. Melting of solid I was
followed up to 808~K and 11.7~GPa. At 810~K and 11.8~GPa the sample suddenly transited into pure phase VII.  It
was then decompressed in order to melt it and heated to 827~K. Increasing the load at this temperature produced
the equilibrium between solid VII and fluid at 12.1~GPa. We then followed the melting curve of phase VII up to
920~K. Compression of phase VII at 879~K produced  phase IV at 16.8~GPa. Surprisingly, the latter could be
decompressed down to the equilibrium between phase IV and the fluid, at 884~K and 13.2~GPa. We then measured
several melting points of phase IV from 810~K to 946~K. As shows Fig. 2, this melting curve is not
distinguishable from the one of phase VII within our experimental uncertainties ($\pm0.2$ GPa and $\pm5$ K at
these $P-T$ conditions).

The fact that the solid phase obtained from the melt above 827~K is phase VII confirms that it is indeed the
stable phase at these temperatures and pressures. However, the closeness of the melting curves shows that the
energy difference between phase IV and VII is very small, since at melting the Gibbs free energy of the fluid
and solid phases are equal.  The fact that we could decompress phase IV down to the solid/fluid equilibrium
without reverting to phase VII gives an indication that there is a strong potential barrier hindering this
transformation. This also explains why we were able to measure the transition line between phase I and IV by
compression of phase I even in the temperature range where we observed phase VII: since we tuned the pressure
closely around the transition, it is likely that part of the sample was still in phase IV, forcing the transition
from phase I to IV without passing via phase VII.

Fig. \ref{Fig3} shows that the VII-IV transition  is accompanied by a small volume decrease ($0.66\%$ at 726~K).
Since the slope of the transition line is positive, the Clapeyron's law imply that phase VII has a larger entropy
(by about $4$~J/mole) than phase IV. This agrees with the calculation of Bonev \textit{et al.}~\cite{BonGyg2003}
who found that the $Cmca$ structure is stabilized at high temperatures by its larger entropy due to the
contribution of a soft acoustic phonon. Extrapolating the measured melting line and VII-IV transition line gives
a possible IV-VII-F triple point  at $\approx$1030~K and 18.4~GPa.

Phase VII is thus the stable solid phase beyond phase I for T$ > 640$~K but phase IV becomes more stable at
higher pressures. This suggests that the structure of phase IV needs to be revised, since the proposed $Pbcn$
structure~\cite{ParYoo2003} was found by Bonev \emph{et al}.~\cite{BonGyg2003} to be thermodynamically unstable
with respect to the $Cmca$ structure in the whole $P-T$ field of phase IV.

Finally, our findings are useful in clarifying the present debate on the molecular character of the high $P$-$T$
phases IV and II of carbon dioxide. As mentioned in the beginning, the C$=$O bond lengths found experimentally
for these two phases \cite{YooKoh2002,ParYoo2003} are much larger than in phase I~\cite{DowSom1998}, which was
taken as evidence that these phases are intermediate forms between the molecular and covalent bonded solids.
Bonev \emph{et al.}~\cite{BonGyg2003} found that the bond lengths in either of phase II or IV structure shortened
to a value comparable to the one in phase I; the calculated energy difference between the long and short bond
lengths is in excess of 3 and 6~eV per molecule respectively in phase II and IV. We have shown here that phase
VII is a strictly molecular phase, with a C$=$O bond length close to that of phase I at 1 GPa.  Moreover, from
the volume and entropy jumps at the VII-IV transition, we estimate an internal energy difference of about 13 meV
per molecule between the two phases at 726 K: this strongly suggests that phase IV, and most likely phase
II~\footnote{Ref.~\onlinecite{BonGyg2003} reports a calculated energy difference of 20.6~meV/molecule between the
molecular $Cmca$ and $P4_2/mnm$ (phase II) structures.}, are strictly molecular phases too, in agreement with a
recent spectroscopic study~\cite{GorGio2004}.

We thank B. Canny (IMPMC) and M. Mezouar (ESRF ID27) for their help with the experiments.

\end{document}